\documentclass[12pt,a4paper,twoside]{amsart}
\usepackage{txfonts}

\usepackage[author-year]{amsrefs}

\usepackage{amsthm}

\usepackage{graphicx}

\usepackage{algorithm}

\usepackage{algorithmic}
\usepackage{amsmath}

\theoremstyle{plain}

\theoremstyle{definition}

\theoremstyle{remark}

\makeatletter

\def\ps@plain{\ps@empty
  \def\@oddfoot{\normalfont\scriptsize \hfil\hfil}%
  \let\@evenfoot\@oddfoot}
\def\ps@headings{\ps@empty
  \def\@evenhead{%
    \setTrue{runhead}%
    \normalfont\large
    \rlap{}\hfil \leftmark{}{}\hfil}%
  \def\@oddhead{%
    \setTrue{runhead}%
    \normalfont\large \hfil
    \rightmark{}{}\hfil \llap{}}%
  \let\@mkboth\markboth
}
\def\ps@firstpage{\ps@plain
  \def\@oddfoot{\normalfont\scriptsize \hfil\hfil
     \global\topskip\normaltopskip}%
  \let\@evenfoot\@oddfoot
  \def\@oddhead{\@serieslogo\hss}%
  \let\@evenhead\@oddhead 
}

\def\@setaddresses{\par
  \nobreak \begingroup
\large
  \def\author##1{\nobreak\addvspace\bigskipamount}%
  \def\\{\unskip, \ignorespaces}%
  \interlinepenalty\@M
  \def\address##1##2{\begingroup
    \par\addvspace\bigskipamount\indent
    \@ifnotempty{##1}{(\ignorespaces##1\unskip) }%
    {\scshape\ignorespaces##2}\par\endgroup}%
  \def\curraddr##1##2{\begingroup
    \@ifnotempty{##2}{\nobreak\indent{\itshape Current address}%
      \@ifnotempty{##1}{, \ignorespaces##1\unskip}\/:\space
      ##2\par}\endgroup}%
  \def\email##1##2{\begingroup
    \@ifnotempty{##2}{\nobreak\indent{\itshape E-mail address}%
      \@ifnotempty{##1}{, \ignorespaces##1\unskip}\/:\space
      \ttfamily##2\par}\endgroup}%
  \def\urladdr##1##2{\begingroup
    \@ifnotempty{##2}{\nobreak\indent{\itshape URL}%
      \@ifnotempty{##1}{, \ignorespaces##1\unskip}\/:\space
      \ttfamily##2\par}\endgroup}%
  \addresses
  \endgroup
}

\def\@maketitle{%
  \let\@makefnmark\relax  \let\@thefnmark\relax
  \ifx\@empty\@date\else \@footnotetext{\@setdate}\fi
  \ifx\@empty\@subjclass\else \@footnotetext{\@setsubjclass}\fi
  \ifx\@empty\@keywords\else \@footnotetext{\@setkeywords}\fi
  \ifx\@empty\thankses\else \@footnotetext{%
    \def\par{\let\par\@par}\@setthanks}\fi
  \@mkboth{\@nx\shortauthors}{\@nx\shorttitle}%
  \global\topskip42\p@\relax 
  \@settitle
  \ifx\@empty\authors \else \@setauthors \fi
  \ifx\@empty\@dedicatory
  \else
    \baselineskip18\p@
    \vtop{\centering{\large\itshape\@dedicatory\@@par}%
      \global\dimen@i\prevdepth}\prevdepth\dimen@i
  \fi
  \@setabstract
 \normalsize
  \if@titlepage
    \newpage
  \else
    \dimen@34\p@ \advance\dimen@-\baselineskip
    \vskip\dimen@\relax
  \fi
}

\long\def\@footnotetext#1{%
  \insert\footins{%
\normalsize
    \interlinepenalty\interfootnotelinepenalty
    \splittopskip\footnotesep \splitmaxdepth \dp\strutbox
    \floatingpenalty\@MM \hsize\columnwidth
    \@parboxrestore \parindent\normalparindent \sloppy
    \protected@edef\@currentlabel{%
      \csname p@footnote\endcsname\@thefnmark}%
    \@makefntext{%
      \rule\z@\footnotesep\ignorespaces#1\unskip\strut\par}}}

  \def\@oddhead{\thepage\hfil\textsc{ Kateryna Mishchenko, Volodymyr
Mishchenko and Anatoliy Malyarenko}\hfil}
  \def\@evenhead{\hfil\textsc{Newton-type Methods for REML Estimation\dots}\hfil\thepage}

\renewenvironment{abstract}{%
  \ifx\maketitle\relax
    \ClassWarning{\@classname}{Abstract should precede
      \protect\maketitle\space in AMS documentclasses; reported}%
  \fi
  \global\setbox\abstractbox=\vtop \bgroup
    \normalfont\normalsize
    \list{}{\labelwidth\z@
      \leftmargin3pc \rightmargin\leftmargin
      \listparindent\normalparindent \itemindent\z@
      \parsep\z@ \@plus\p@
      
    }%
    \item[\hskip\labelsep\scshape\normalsize\abstractname.]%
}{%
  \endlist\egroup
  \ifx\@setabstract\relax \@setabstracta \fi
}

\def\@captionfont{\large\normalfont}

\makeatother

\begin{document} {\large

{\small\hfill Research Reports MdH/IMa}

{\small\hfill No. 2007-8, ISSN 1404-4978}

\vspace{15mm} \addtolength{\topmargin}{-0.7cm}

\title{Newton-type Methods for REML Estimation in Genetic Analysis
of Quantitative Traits}

\author[]{Kateryna Mishchenko}
\address{Department of Mathematics and
Physics, M\"{a}lardalen University, Box 883, SE-721 23
V\"{a}ster{\aa}s, Sweden}
 \email{kateryna.mishchenko@mdh.se}

\author[]{Sverker Holmgren}
\address{Division of Scientific Computing, Department of
Information Technology, Uppsala University, Sweden}
 \email{sverker@it.uu.se}
\author[]{Lars R\"onneg{\aa}rd}
\address{Linn{\ae}us Centre for Bioinformatics, Uppsala University, Sweden}
 \email{Lars.Ronnegard@lcb.uu.se}
\date{\today}

\begin{abstract}
Robust and efficient optimization methods for variance component
estimation using Restricted Maximum Likelihood (REML) models for
genetic mapping of quantitative traits are considered. We show that
the standard Newton-AI scheme may fail when the optimum is located
at one of the constraint boundaries, and we introduce different
approaches to remedy this by taking the constraints into account. We
approximate the Hessian of the objective function using the average
information matrix and also by using an inverse BFGS formula. The
robustness and efficiency is evaluated for problems derived from two
experimental data from the same animal populations.
\end{abstract}
\maketitle

\section{Introduction}

One of the goals in modern computational genetics is to locate
regions in the genome underlying quantitative traits. This is
performed by mapping of quantitative trait loci (QTL), which is a
procedure involving statistical analysis of data sets derived from
experimental populations. QTL mapping is based on the idea of relating
phenotypic and marker genotype information.
The QTL are regions on the genome where the genetic marker information
and the phenotypic values show strong co-variation.
We focus on experimental crosses where animals from two divergent breeds
have been mated for two generations producing
a large number of grand-offspring.

In its simplest form, QTL analysis assumes that all animals within
the two divergent breeds show no genetic variation with the founder
breeds \cite{HK92}, which is motivated by the expectation of having
most of the genetic variation between breeds. There may be
substantial genetic variation within breeds, which may be taken into
account in a variance component QTL model \cite{FeGr89,Go90,RC07}.
This is a mixed linear model with fixed non-QTL effects and a random
QTL effects. Here, either Maximum Likelihood (ML) or Restricted
Maximum Likelihood (REML) methods are used and the QTL are given by
the regions having highest likelihood ratio statistic.

ML maximizes parameter values for both fixed and random effects
simultaneously, whereas REML maximizes only the portion of the
likelihood that does not depend on fixed effects.  An introduction
to ML and REML schemes applied to QTL mapping problems is given e.g.
in \cite{LyWa97} and Chapter 23 of \cite{Meychap23}.

In this paper we develop and assess efficient and robust
computational procedures for solving REML estimation problems in QTL
mapping settings where variance component models are used.

\section{Linear Mixed Models and REML Estimation}
A general linear mixed model is given by
\begin{equation}
y = Xb + Zu + e \label{mme}
\end{equation}
where $y$ is a vector of $n$ observations, $X$ is the $n\times n_f$
design matrix for $n_f$ fixed effects, $Z$ is the $n \times n_r$
design matrix for $ n_u$ for random effects, $b$ is the vector of
$n_{f}$ unknown fixed effects, $u$ is the vector of $n_u$ unknown
random effects, and $e$ is a vector of n residuals of random
effects.

The additional assumptions for the QTL analysis setting are that
elements of $e$ are identically and independently distributed and
that there is a single observation for each individual. In this
paper, we focus on the case where the model includes a single random
effect. The covariance matrix for (\ref{mme}) then becomes
\begin{equation}
V = \sigma^{2}_{\pi} \Pi+ \sigma^{2}_{e}I\label{matV},
\end{equation}
where $\sigma^{2}_{\pi}$ is the variance of the random effect and
$\sigma^{2}_{e}$ is the residual variance. The matrix $\Pi$ is
referred to as the {\em Identity-By-Descent (IBD)} matrix.

In REML estimation, the task at hand is to determine estimates of
$\sigma^{2}_{\pi}$ and $\sigma^{2}_{e}$ in (\ref{mme}) as well as
the value of the likelihood function $l$ at these points. At least
two approaches can be used for this. One standard scheme is based on
computing the estimates from the Mixed-Model Equations (MME)
introduced in \cite{He63}. However, this approach requires that the
IBD matrix is sparse and nonsingular, which is normally not the case
in the QTL analysis problems. Instead, we use an alternative
approach which is also used in the standard software available for
REML models for QTL mapping problems. Comparison of these two
approaches is given in \cite{LeWe06}. Here,

the parameters $\sigma^{2}_{\pi},\sigma^{2}_{e}$ are obtained as
maximizers of the restricted likelihood of the observed data $y$.

The log-likelihood function for the REML estimation approach based
on model (\ref{mme}) is
\begin{equation}
L \equiv -2ln(l) = C + ln(det(V))+ ln(det(X^{T}V^{-1}X))+ yP^{T}y,
\label{llh}
\end{equation}
where $l$ is the likelihood function to be maximized and the projection matrix $P$ is defined by
\begin{equation}
P =V^{-1} - V^{-1}X(X^{T}V^{-1}X)^{-1}X^{T}V^{-1}.\label{matP}
\end{equation}
The function $L$ is a function of two variables,
$$ L(\Sigma) = L(\sigma^{2}_{\pi},\sigma^{2}_{e}) \equiv
L(\sigma_{1},\sigma_{2}),$$ and the problem of maximizing the
likelihood function $l$ is equivalent to the problem of the
minimizing the log-likelihood function $L$, which has a simpler
representation. In summary, we determine the estimates of
$\sigma^{2}_{\pi},\sigma^{2}_{e}$ by solving the optimization
problem
\begin{eqnarray}
min\ \ L(\sigma_1,\sigma_2) \label{min} \\
s.t. \ \ \sigma_{1}\ge 0 \label{con1}\\
\sigma_{2}> 0 \label{con2}
\end{eqnarray}

The generalized least square estimates of the fixed effect $b$ may
be computed by first solving the optimization problem (\ref{min}) -
(\ref{con2})
 and then computing $b$ using
\begin{equation}
\widehat{b} = (X^{T}V^{-1}X)^{-1}X^{T}V^{-1}y. \label{matb}
\end{equation}
  where
$\sigma_{1},\sigma_{2}$ in $V$ are the optimizers for the problem
(\ref{min}) - (\ref{con2}).


\section{A Brief Review of Optimization Methods for
Maximum Likelihood Computations} Several methods have been used for
solving the optimization problems arising from maximum-likelihood
estimation schemes. The algorithms can be classified in several
groups, e.g. derivative-free, derivative-based (Newton-like), and
expectation-maximization (EM) methods, method of successive
approximations (MSA). These schemes can also be combined in
different ways. A review is given in
\cite{DrDu06,Ha77,Meychap23,LyWa97}.

The choice of method for maximizing the likelihood function is
affected mainly by two factors: Firstly, the computation of the
likelihood and its derivatives is computationally demanding, which
means that the optimization algorithm should require a small amount
of such evaluations in each step. Secondly, the optimization
algorithm should be efficient and robust, which means that it should
converge in a small number of iterations and the performance should
not depend critically on the initial values and the properties of
the objective function for the specific problem.
\subsection{Derivative-free methods}
Derivative-free schemes only employ evaluations of the objective
function. Examples are the Nelder-Mead downhill simplex method
\cite{NeMe64} and the quasi-Newton method with finite-difference
approximation of the derivatives. For REML problems, evaluation of
the log-likelihood is rather costly, and the Quasi-Newton method
with finite-difference approximation is not a very efficient scheme.
However, it has been shown that the Quasi-Newton method has better
convergence properties than the downhill simplex method, and the
latter method can also exhibit non-robust behavior when the initial
point is close to the maximum \cite{Me89}.
\subsection{Methods using derivative information}
The standard schemes for optimization in REML schemes use derivative
information. These methods are based on solution of the nonlinear
equation
\begin{equation}
 DL(\Sigma) = 0, \label{DLeq}
\end{equation}
where $DL = (\frac{\partial L}{\partial \sigma_{1}},\frac{\partial
L}{\partial \sigma_{2}})^{T}$ is the gradient vector of the
log-likelihood function. The components of the gradient are
expressed in terms of the matrices $V$ and $P$ and the variance
component parameters $\sigma_{1,2}$ using
\begin{equation}
\frac{\partial L}{\partial \sigma_{i}} = tr(\frac{\partial
V}{\partial \sigma_{i}}P) - y^{T}P\frac{\partial V}{\partial
\sigma_{i}}Py\,,\quad \ i = 1,2. \label{DL}
\end{equation}
For REML estimation problems, it has been shown that Newton-type
methods are quite efficient \cite{CaHa91,Ha77,JoTho95}. The EM
method introduced in \cite{DeLaRu77}, which is often used in general
maximum likelihood settings, is not guaranteed to convergence to the
true minimum and requires more iterations than Newton-type methods
\cite{LyWa97}. However, the modifications of this method were shown
to be efficient, see e.g. \cite{CaHa91,ThoMe86}.

The standard Newton method is defined by the iteration
\begin{equation}
\Sigma^{k+1} = \Sigma^{k} - \alpha^k
[H(\Sigma^{k})]^{-1}[DL(\Sigma^{k})], \label{newton}
\end{equation}
with $\alpha^k = 1$ and  $H(\Sigma^k)\equiv H^k$ is the Hessian of
the log-likelihood function, which for a REML problem is given by
\begin{eqnarray}
H(\sigma_{i},\sigma_{j}) 
= - tr(\frac{\partial V}{\partial \sigma_{i}}P\frac{\partial
V}{\partial \sigma_{j}}P) +2y^{T}P\frac{\partial V}{\partial
\sigma_{i}}P\frac{\partial V}{\partial \sigma_{j}}Py\, ,\quad \ i,j
=1,2. \label{hessian}
\end{eqnarray}
The true Hessian is expensive to evaluate (especially the first
term), and two approximations have been used:
\begin{enumerate}
\item \textbf{Fisher's method of scoring:} The
Hessian in (\ref{newton}) is substituted by it's expected value:
$H^{k}\longrightarrow E(H^{k}) = -F^{k}$, where $F^{k}$ is the
Fisher information matrix, see e.g. \cite{Tho73}.

\item \textbf{Average Information (AI) method:}
The Newton-AI scheme is a standard method for solving REML problems,
used e.g. in \cite{JoTho95}. Here, the Hessian in (\ref{newton}) is
substituted by the so called average information matrix (see
\cite{GiThoCu95}): $H^{k}\longrightarrow
\frac{1}{2}(H^{k}+E(H^{k}))= AI^k$. The AI matrix is given by
\begin{equation}
AI(\sigma_i,\sigma_j) = \frac{1}{2}(\frac{\partial^{2} L}{\partial
\sigma_i
\partial \sigma_j} + E(\frac{\partial^{2} L}{\partial \sigma_i
\partial \sigma_j})) = y^{T}P\frac{\partial V}{\partial \sigma_i}P\frac{\partial V}{\partial
\sigma_j}Py\, ,\quad \ i,j =1,2.\label{matrai}
\end{equation}
\end{enumerate}
As an initial step in this study, the Newton-AI method, as described
in \cite{JoTho95}, was implemented and tested. The results show that
this method results in good performance for cases where the maximum
is inside the region restricted by the constraints. However, for
problems where the maximum is at or close to the constraints, the
results presented in Section \ref{secnumstqn} show that the method
may break down since the constraints are violated.

In general, (\ref{min}) - (\ref{con2}) presumably should be solved
using some optimization method that takes the non-negativity
constraints for the variance component parameters into account, see
e.g. \cite{CaHa91,Ha77,MeSm96}. Also, the optimization problem may
in some cases be non-convex in parts of the domain and/or the
objective function may be very flat in one direction. In the next
section, we present some methods which take these properties of the
optimization problem into account.

\section{New Optimization Procedures for REML Estimation}

We present three different methods: The standard Newton-AI method
enhanced with a line search scheme that takes the constraints into
account. A quasi-Newton scheme where the same type of line search
scheme is included and the scheme is also modified to deal with
non-convex parts of the objective function, and finally an
active-set method where the treatment of the constraints is built
into the algorithm.

\subsection{An enhanced Newton-AI algorithm}\label{secnewai}
Standard unconstrained optimization methods can be modified to take
simple constraints into account by introducing a line search scheme
which prevents constraint violation, and has been suggested for application
in REML estimation \cite{Ha77,Je97}. We use a simple line search
procedure where the conditions (\ref{con1}) and (\ref{con2}) are
initially checked for the full-length step $\alpha^k = 1$, and if
the constraints are not fulfilled the step length is reduced by a
factor of two. Then the constraints are checked again and the step
length is reduced further if needed. The Newton iteration is
terminated if the relative step length $\alpha^k$ is smaller than
some pre-set parameter, e.g. $10^{-5}$. This means that if an
optimum on the boundary is found, the line search plays a key role
and the line search termination criterion stops the iteration.

Generally, the line search procedure maybe used as well to avoid the
"overshooting" the minimum or to enforce a decrease of the
log-likelihood function.  To this purpose the described above line
search technique was used e.g. in \cite{JeSa76}. Moreover,
 there are more
efficient line search techniques such as Armijo or Wolf line
searches. The line searches aimed to reducing the value of the
objective function require additional function evaluations, which is
very computationally demanding, so undesirable. Moreover, in our
algorithm we skip the direct function evaluations. That is why, in
our study we use the line search only for feasibility checking,
since this does not require any additional function evaluation.

\subsection{An Enhanced BFGS-Quasi-Newton Method}\label{secbfgsnew}
An obvious candidate for solving the REML optimization problem is
the quasi-Newton (QN) method, where the Hessian is adaptively
approximated instead of using e.g. the average information matrix as
in the Newton-AI scheme. The QN-iteration is given by
\begin{equation}  \label{qn}
\left.
\begin{aligned}
  p^{k+1} & = - [\tilde
H(\Sigma^{k})]^{-1}[DL(\Sigma^{k})]\\
  \Sigma^{k+1} & = \Sigma^{k} + \alpha^k\cdot p^{k+1}
\end{aligned}
 \right\}
\end{equation}

where $\tilde H(\Sigma^k)\equiv \tilde H^k$ is the approximation of
the Hessian at iteration $k$. A reason for using this method is that
it is cheaper to update the approximation of the Hessian than to
compute the AI matrix.

There are several updating formulas available for the approximative
Hessian in a QN scheme, see e.g. \cite{Nocedal}. We use a inverse
BFGS (Broyden-Fletcher-Goldfarb-Shanno) formula, where an
approximation of the inverse of the Hessian $B^k \equiv (H^k)^{-1}$
is updated in each iteration. In the QN scheme, the gradient is
calculated using (\ref{DL}). Using a finite difference approximation
would be inefficient, since it requires evaluation of the
log-likelihood which is computationally expensive.

The inverse BFGS formula produces a positive definite Hessian
approximation matrix if the curvature condition
\begin{equation}
(\Delta\Sigma^{k})^{T}\Delta (D L^{k}) > 0 \label{curcon}\
\end{equation}
holds. Here $\Delta (D L^{k}) = (D L)^{k+1}-(D L)^{k}$. However,
this condition does not hold if the objective function is
non-convex. In this case (\ref{curcon}) can be enforced explicitly,
by imposing restrictions on the step length in the line search
procedure, see \cite{Nocedal}. For the REML optimization problems,
the recursive reduction of the step-length in the standard line
search algorithms such as e.g. Armijo line search should be avoided
since they involve evaluations of the log-likelihood. A standard
approach to avoid problems caused by an indefinite approximation of
inverse Hessian
 is to skip the updating and use
\begin{equation}
\tilde B^{k+1} = \tilde B^{k} \label{updhes}
\end{equation}
when (\ref{curcon}) is not fulfilled. However, we found that for the
QTL mapping problems this type of algorithm sometimes failed since
using (\ref{updhes}) ignores a lot of information about the real
curvature of the function. Instead, we propose to use the inverse AI
matrix as the next approximation of the inverse of the Hessian if
condition (\ref{curcon}) is not satisfied. The results in Section
\ref{secnumex} show that this gives an efficient algorithm. Also,
the performance of the QN scheme is enhanced if the initial value of
the inverse Hessian matrix is set to the inverse of the AI matrix.

The modified inverse BFGS procedure is given by the following
algorithm:
\begin{algorithm}[H]
\caption{Modified inverse BFGS updating formula}\label{alg1}
\begin{algorithmic}
 \STATE $s = \Sigma^{k+1} - \Sigma^{k}$
 \STATE $y = \nabla L^{k+1}-\nabla L^{k}$
 \IF{(k=0) OR ($s^T\cdot y \le 0$)}
\STATE
\begin{equation}
\tilde B^{k} = AI^{-1}
\end{equation}
\ELSE \STATE $\rho=\frac{1}{y^{T}\cdot s}$
\begin{equation}
\tilde B^{k}  = (I-\rho\cdot s\cdot y^{T})\cdot \tilde B^{k-1}\cdot
(I-\rho\cdot y\cdot s^{T})+\rho\cdot s\cdot s^{T}
\end{equation}
\ENDIF
\end{algorithmic}
\end{algorithm}
The line search procedure described above, ensuring that the
constraints are fulfilled, is used in the same way in the QN method
as in Newton-AI method.
\subsection{An Active-Set Method}\label{secai}
\par The active-set method is a Newton-type method for
constraint problems, see e.g. \cite{NaSo96}.
 The constraints are automatically satisfied in
each iteration, and the gradient and Hessian are calculated in a
reduced space: If $A$ is the matrix of gradients of active
constraints and $N$ is the null space of the matrix $A$, the reduced
gradient and Hessian are given by $N^{T}DL$ and $N^{T}HN$. The
iterative scheme for this method is:
\begin{equation}\label{as}
\left.
\begin{aligned}
  p^{k+1} & = -N[N^{T}\cdot H(\Sigma^k)\cdot
 N]^{-1} \cdot N^{T}DL(\Sigma^k) \\
  \Sigma^{k+1} & = \Sigma^{k} + \alpha^k\cdot p^{k+1}
\end{aligned}
 \right\}
\end{equation}
 The optimality condition is checked by examining of Lagrangian
multipliers $\lambda$ at the point of potential optimum $\Sigma^*$:
\begin{equation}
\lambda = A_{r}^T DL(\Sigma^*)\label{lambda}
\end{equation}
where $A_{r}$ is right inverse of matrix $A$ computed as
\begin{equation}\label{matrA}
A_{r} = A^T [A A^T]^{-1}
\end{equation}

The active-set strategy for the REML was implemented in
\cite{CaHa91} as well. In our study we approximate the Hessian both
using the AI matrix and the inverse BFGS formula (the Hessian $H^k$
is approximated by $(\tilde B^k)^{-1}$). In this case we use a line
search procedure where the current step is, if needed, reduced so
that the next point lies exactly on the relevant constraint. The
step length $\alpha^k$ is controlled by the pre-set parameter
$10^{-5}$ and iterations terminate if the current step length is
smaller than this value.

If no constraints are active, the active-set and corresponding
unconstrained Newton methods (AI and QN) are equivalent, and the
same sequence of iterations are generated.

 The numerical procedure for all methods
are summarized in algorithm described in the  Algorithm \ref{alg2}.
\begin{algorithm}[]
\caption{Short Algorithm of AS and QN Methods}\label{alg2}
\begin{algorithmic}

 \STATE \texttt{Initialization}:

 \STATE Set up A, X, y, $\Sigma^0$, type of Hessian approximation
  \STATE resvar = var$(y^{T}-X\cdot(X^{T}\cdot X)^{-1}\cdot X^{T}\cdot
y^{T})$
 \STATE $\Sigma^{1} = \Sigma^0 \cdot$ resvar
 \STATE $k = 0$

 \STATE \texttt{Main loop}:
 \WHILE{ $\|DL\|_{2}^{2} \ge 10^{-6}$}
    \IF{((AS) and ($\|N^{T}\cdot DL\|_{2}^{2} \leq 10^{-6}$))}

       \IF {(no active constraints)} \STATE \textbf{break}; \ENDIF
       \STATE Compute $\lambda$ by (\ref{lambda})
       \IF {($\lambda \geq 0$)} \STATE \textbf{break}; \ENDIF
       \STATE update $A$ and $N$

    \ENDIF

   \STATE $k = k+1$
   \STATE Compute $V$ by (\ref{matV})

   \STATE Compute $P$ by (\ref{matP})

   \STATE Compute DL by (\ref{DLeq})

   \STATE Compute Hessian depending on type of Hessian approximation:

   \STATE ($\tilde{B}^k$ by Algorithm \ref{alg1}), $\tilde{H}^k$ or ($ AI^{-1}$ by (\ref{matrai}))

   \STATE \texttt{Newton-AI or Active-Set Method Step}

   \STATE Compute direction $p^{k}$ from (\ref{qn}) or (\ref{as})
   \STATE Find step length $\alpha^k$:
   \IF {($\alpha < 10^{-5}$)} \STATE \textbf{break}; \ENDIF
   \STATE $\Sigma^{k+1} = \Sigma^{k} + \alpha^{k} \cdot p^{k+1}$
   \IF {((AS) and (new active constraints))} \STATE update $A$ and $N$;\ENDIF

\ENDWHILE

\STATE Evaluate $\beta$ by (\ref{matb})

\end{algorithmic}
\end{algorithm}

\section{\textbf{Numerical Experiments}}\label{secnumex}

In this section, we present numerical experiments performed for two
representative sets of experimental data that come from the same
population. We compute the optimal values of the variance of a
single random effect ($n_u=1$) and the residual, i.e. we solve
two-dimensional optimization problems for $\sigma_{1,2}$ under the
constraints that $\sigma_{1}\ge 0, \sigma_2>0$. The population size
is $n=767$ (which is quite typical in QTL analysis) and there is a
single fixed effect ($n_f=1$). For data set 1, the optimum is
distinctly defined and located inside the feasible search domain,
the optimal values are $\sigma_1 = 4868$ and $\sigma_2 = 20644$. For
data set 2, the optimum is found at one of the constraint
boundaries, the optimal values are $\sigma_1 = 0$ and $\sigma_2 =
29681 $. Also, the objective function is non-convex and very flat at
the optimum in the $\sigma_2$-direction, making the computed optimal
value of $\sigma_2$ very sensitive.

In practice the log-likelihood is evaluated with the accuracy of
$4-5$ decimals. In our implementation we skipped the evaluation of
the log-likelihood to make our computations cheaper.  As a
termination criterium the magnitude of the gradient/reduced gradient
were used, these quantities are computed by as a part of the
algorithm, so they do not require extra computational work.

\subsection{The Standard Newton-AI Method}\label{secnumstqn}
We begin by showing results for a case where the standard Newton-AI
method fails. For this method, we use the termination criterion $\|
DL\|_{2}^{2} \le 10^{-6}$. In Figure \ref{f2}, the values of the
objective function, the norm of gradient, and the variance
components are shown as functions of the iteration number in the
Newton scheme for data set 2. For this problem, the optimum is found
at the constraint $\sigma_1=0$. From Figure \ref{f2}, it is clear
that the constraint is violated and $\sigma_{1}$ becomes negative
already after the first iteration.
\begin{figure}[tbp]
\centering
\includegraphics[width=4in]{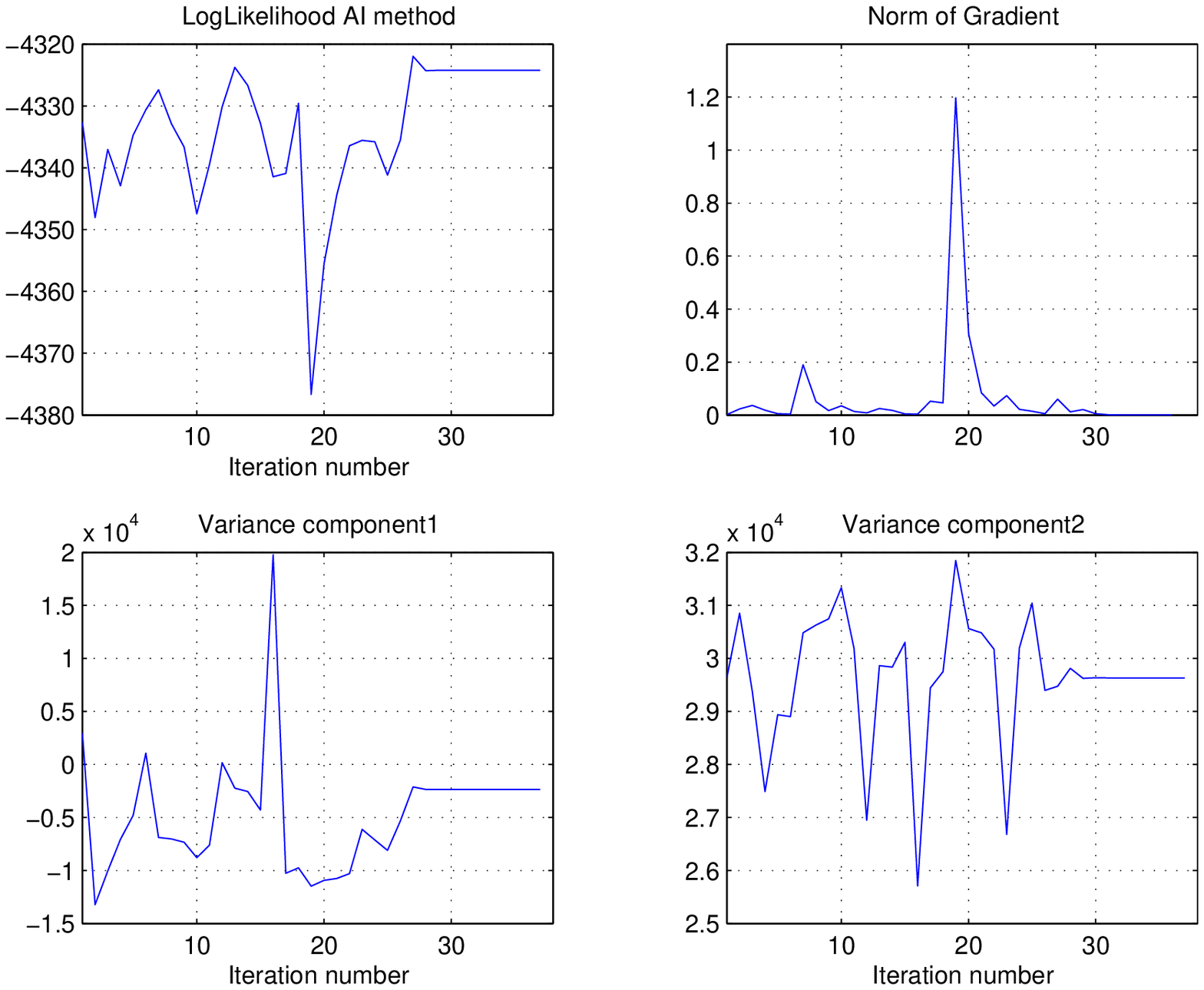}
\caption{The standard Newton-AI method for data set 2}\label{f2}
\end{figure}

\subsection{The Enhanced Newton-AI Method}
\begin{figure}[tbp]
\centering
\includegraphics[width=4in]{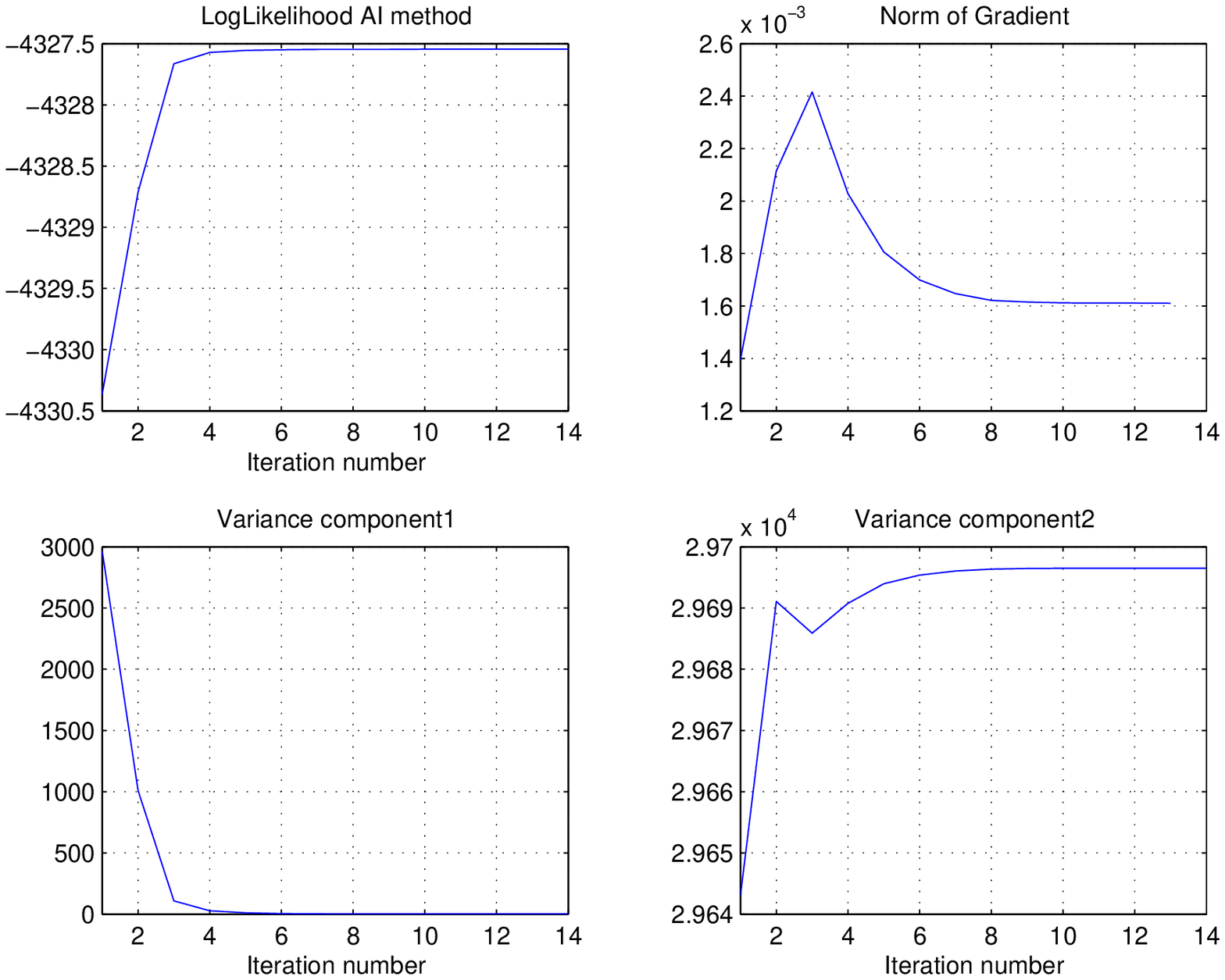}
\caption{The enhanced Newton-AI method, including line search, for
data set 2}\label{f3}
\end{figure}
In Figure \ref{f3}, the results for the enhanced Newton-AI method
described in Section \ref{secnewai} are shown for data set 2.
 In
this case, the termination criterion $\alpha^{k} \le 10^{-5}$ for
the line search is added. This is also the criterion responsible for
stopping the iteration, and from Figure \ref{f3} it is clear that
the enhanced method does find the correct minimum, despite the fact
that the objective is non-convex. Introducing the line search
procedure does not change the performance of the method for
unconstrained problems, i.e. for data set 1. In this case, the
standard Newton-AI scheme and the enhanced version produce the same
iteration sequences and the same (quite acuurate) results.

\subsection*{The Enhanced Quasi-Newton method}
We now present results for the enhanced quasi-Newton method for data
sets 1 and 2, and compare these results to those of the enhanced
Newton-AI scheme. In Tables \ref{tab2} and \ref{tab4}, the
convergence histories for the two data sets are compared.

\begin{table}[tbpf]
\begin{center}

\begin{tabular}{|c|r|c|r|c|}
 \hline
& \multicolumn{2}{|c|}{} & \multicolumn{2}{|c|}{ }\\
& \multicolumn{2}{|c|}{Quasi-Newton} & \multicolumn{2}{|c|}{Newton-AI}\\
\cline{2-5}
 &                & & & \\
iter.& $\sigma_1$, $\sigma_2$ \ \ \ \ \ \ \ \ & L($\sigma_1$, $\sigma_2$) & $\sigma_1$, $\sigma_2$ \ \ \ \ \ \ \ \ & L($\sigma_1$, $\sigma_2$)\\

  &                & & & \\
  \hline

1&2964.3089,  29643.089 & -4239.3313& 2964.3089,  29643.089 & -4239.3313\\
2& 4810.993,  17002.992 & -4225.8047& 4810.993,  17002.992  &-4225.8047\\
 3&4615.991,  23783.778 & -4222.1107&4761.311,  20046.156 & -4218.9801\\
 4&4749.718,  21684.535  &-4219.2220&4860.177,  20628.874  &-4218.8175 \\
 5&4855.487,  20316.669  &-4218.8628&4869.628,  20644.358 & -4218.8173\\
6& 4845.639,  20679.226  &-4218.8178& 4868.546,  20644.651 & -4218.8173\\
 7&4856.552,  20647.177  &-4218.8174& &\\
8& 4862.895,  20644.420  &-4218.8173&&\\
 9&4868.459,  20644.221  &-4218.8173&&\\
 10&4868.738,  20644.552  &-4218.8173&&\\

  \hline
\end{tabular}
\end{center}
\caption{Convergence histories for data set 1, enhanced quasi-Newton
and Newton-AI schemes}\label{tab2}
\end{table}

\begin{figure}[tbpf]
\centering
\includegraphics[width=4in]{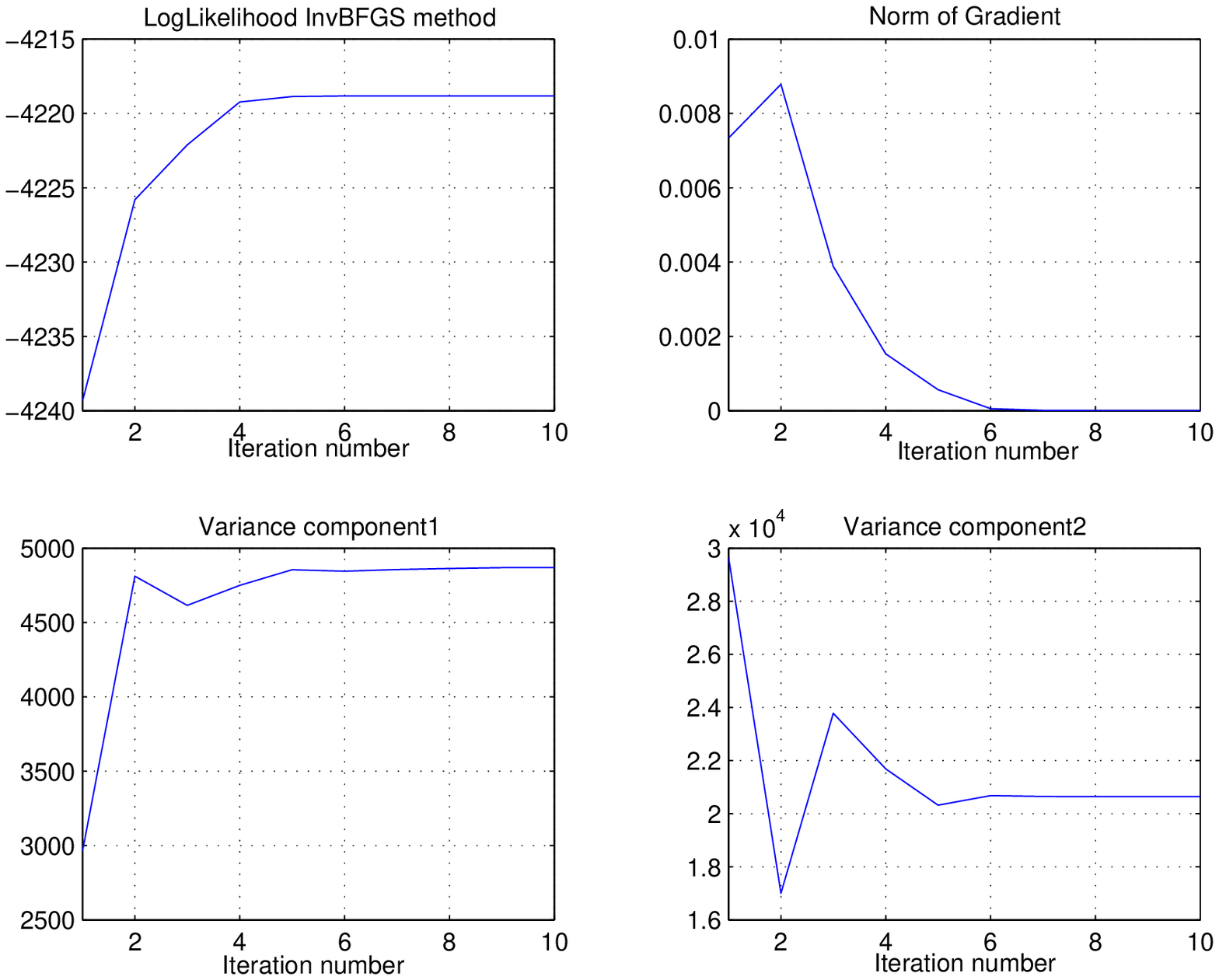}
\caption{The enhanced quasi-Newton method, including line search,
for data set 1}\label{f4}
\end{figure}

\begin{figure}[tbpf]
\centering
\includegraphics[width=4in]{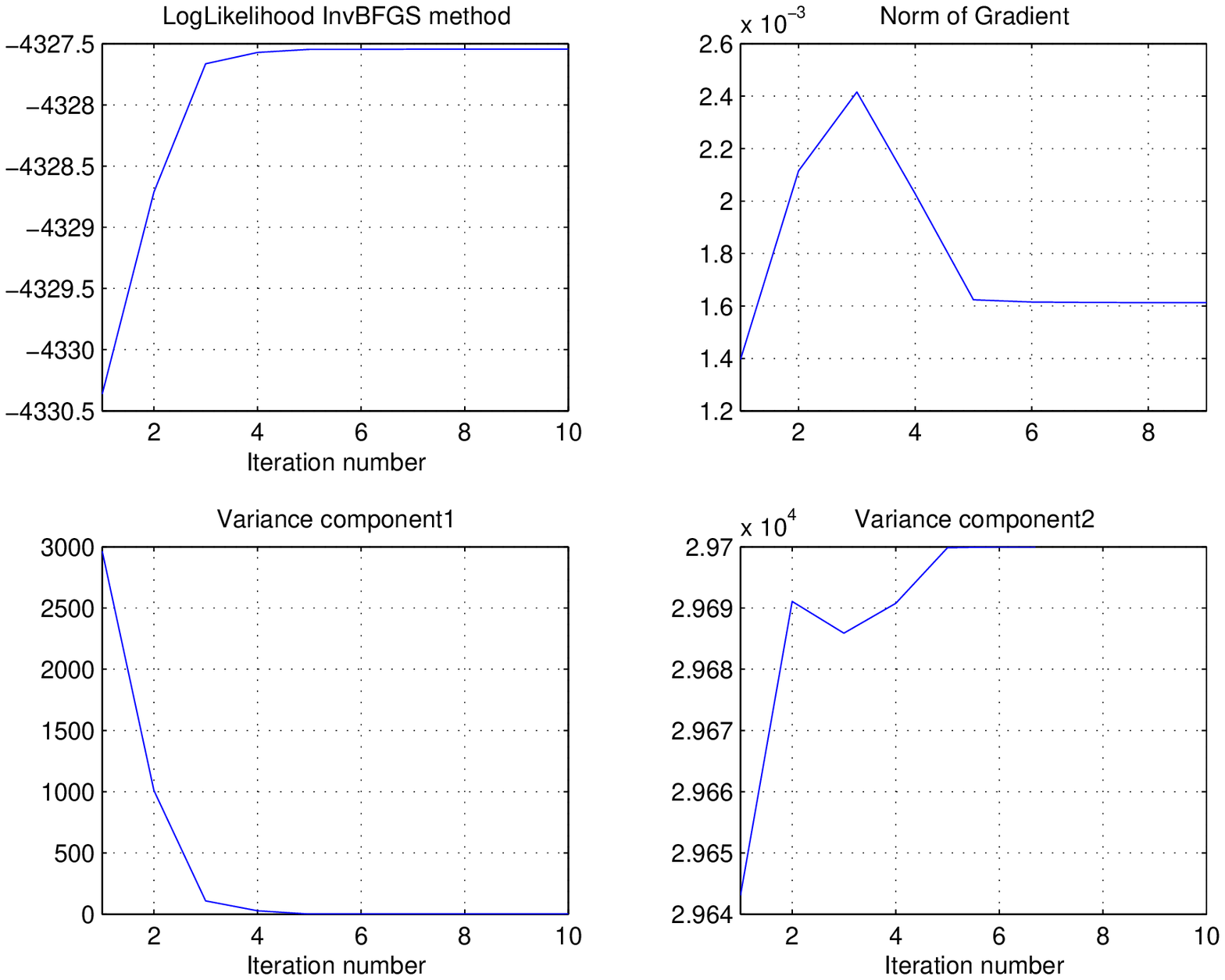}
\caption{The enhanced quasi-Newton method, including line search,
for data set 2}\label{f5}
\end{figure}

\begin{table}[tbpf]
\begin{center}

\begin{tabular}{|c|r|c|r|c|}
 \hline
& \multicolumn{2}{|c|}{} & \multicolumn{2}{|c|}{ }\\
& \multicolumn{2}{|c|}{Quasi-Newton} & \multicolumn{2}{|c|}{Newton-AI}\\
\cline{2-5}
 &                & & & \\
iter.& $\sigma_1$, $\sigma_2$\ \ \ \ \ \ \ \ & L($\sigma_1$, $\sigma_2$) & $\sigma_1$, $\sigma_2$\ \ \ \ \ \ \ \ & L($\sigma_1$, $\sigma_2$)\\

  &                & & & \\
  \hline
1&  2964.3089,  29643.089&  -4330.3621&2964.3089 , 29643.089 &-4330.3621\\
2& 1009.792,  29691.069 & -4328.7130&1009.792,  29691.069 & -4328.7130\\
3& 109.098,  29685.878  &-4327.6635&109.098,  29685.878 & -4327.6635\\
4& 27.693,  29690.757 & -4327.5708&27.693,  29690.757 & -4327.5708\\
5& 0.552,  29699.832 & -4327.5458&10.663,  29693.972 & -4327.5545\\
6& 0.125,  29699.982 & -4327.5455&4.466,  29695.385 & -4327.5490\\
7& 0.049,  29700.008 & -4327.5454&1.777,  29696.043 & -4327.5468\\
8& 0.011,  29700.021 & -4327.5454&0.520,  29696.359 & -4327.5458\\
9& 0.002,  29700.024 & -4327.5454&0.216,  29696.437 & -4327.5455\\
10& 0.001,  29700.024 & -4327.5454&0.065,  29696.476  &-4327.5454\\
 11 &&&0.028,  29696.485  &-4327.5454\\
12&& &0.009,  29696.490  &-4327.5454\\
13 &&&0.004,  29696.491  &-4327.5454\\
14 &&&0.002,  29696.492  &-4327.5454\\

  \hline
\end{tabular}
\end{center}
\caption{Convergence histories for data set 2, enhanced quasi-Newton
and Newton-AI schemes}\label{tab4}
\end{table}

From Table \ref{tab2}, we draw the conclusion that for data set 1,
where the minimum is clearly defined and inside the search region,
the Newton-AI method converges faster that the quasi-Newton scheme.
In this case, the average information matrix is a good approximation
of the Hessian, and inverse BFGS updating formula provides no
improvement.

 The results in Table \ref{tab4} show that for data set
2, the convergence rates of the two methods are different, too. In
this case, the convergence properties are determined not only by the
line search procedure which is the same for the two schemes, but
also by the method of approximation to the Hessian. We note that the
quasi-Newton scheme actually produces faster solution. The first
three iterations of both methods are identical since the objective
function is non-convex in the large region covering the initial
guess. According to the Algorithm \ref{alg1} the quasi-Newton scheme
exploits the average information matrix as Hessian approximation.
After the third iteration the curvature condition (\ref{curcon})
holds, so that the in quasi-Newton method the inverse BFGS
approximation to the Hessian is used. As a result, this method
produces the longer step toward minimum which causes the faster
convergence of the quasi-Newton scheme than the Newton-AI scheme.
Since the objective function is extremely flat at the solution in
the $\sigma_2$-direction, the accuracy of both solutions is similar
and is accepted as a correct one, while the exact solution
($\sigma_1 = 0$) is produced only by the active-set methods.

Generally, the convergence of the unconstrained methods applied to
the problems with optimum at constraint is quite slow in the
neighborhood of the solution which is due to the inefficiency of the
line search procedure. In Figures \ref{f4} and \ref{f5} we show the
results for the quasi-Newton scheme in graphical form.

\subsection{The Active-Set Method}
Finally, we present experiments where the active-set method
described in Section \ref{secai} is used. Here, we use the
termination criterion $\|N^{T}DL\|_{2}^{2} \le 10^{-6}$,
additionally the criterion $\alpha^{k} \le 10^{-5}$ for the line
search is used. As it was for unconstrained methods, this criterion
is responsible for stopping the iterations. Figure \ref{f6} and
\ref{f7} show the convergence history for the active set method
using the average information approximation to the Hessian for data
set 1.

In this case, the optimum is located inside the search region. We
have verified that the results in Figure \ref{f6} are identical to
the corresponding results for the standard Newton-AI scheme, as
expected. When the quasi-Newton scheme with the BFGS formula is used
in the active-set scheme for data set 1, the convergence history is
slightly different from the enhanced quasi-Newton scheme described
above, see Figures \ref{f6} and \ref{f4} . This difference is due to
the way of the approximation to the Hessian: in one case the inverse
of the Hessian is approximated, while in the other case the inverse
of the inverse of the Hessian is computed.

For the active-set method it is interesting to study the performance
for data set 2, where the optimum is located at one of the
constraints. Since the objective function is non-convex, the average
information matrix is used as approximation to the Hessian. As a
result, the methods with the average information and quasi-Newton
approximations of the Hessian produce the identical solutions.
Figure \ref{f8} shows the result for the active-set scheme where the
average information of the Hessian are used. This results should be
compared to Figures \ref{f3} and \ref{f5}, where the corresponding
unconstrained schemes enhanced with a line search procedure are
used. The convergence histories for the active-set methods are shown
in Table \ref{tab8}.

It is clear that the active-set methods produces faster convergence.
The step lengths is each iteration are larger, and the
approximations change more rapidly in the first iterations. In
practice, the minimum is reached after 3-4 iterations. Moreover, the
active-set methods produce more accurate solution than the
unconstrained methods. The general conclusion is that the active-set
approach using the average information approximation for the Hessian
is the most robust scheme. \clearpage
\begin{table}[ht]
\begin{center}
\begin{tabular}{|c|c|c|c|c|}
 \hline
& \multicolumn{2}{|c|}{} & \multicolumn{2}{|c|}{ }\\
& \multicolumn{2}{|c|}{Quasi-Newton} & \multicolumn{2}{|c|}{AI}\\
\cline{2-5}
 &                & & & \\
iter.& $\sigma_1$, $\sigma_2$ & L($\sigma_1$, $\sigma_2$) & $\sigma_1$, $\sigma_2$ & L($\sigma_1$, $\sigma_2$)\\

  &                & & & \\
  \hline
1& 2964.3089,  29643.089&  -4330.3621&2964.3089,  29643.089&-4330.3621\\
2& 0.000,  29715.858  &-4327.5456&0.000,  29715.858  &-4327.5456\\
3& 0.000,  29681.799 & -4327.5453& 0.000,  29681.799 & -4327.5453\\
4& 0.000,  29681.799 & -4327.5453&0.000,  29681.799  &-4327.5453\\

  \hline
\end{tabular}
\end{center}
\caption{Convergence histories for the active-set method using
quasi-Newtion and average information approximation for the Hessian,
data set 2}\label{tab8}
\end{table}
\begin{figure}[]
\centering
\begin{tabular}{c}
\includegraphics[width=4in]{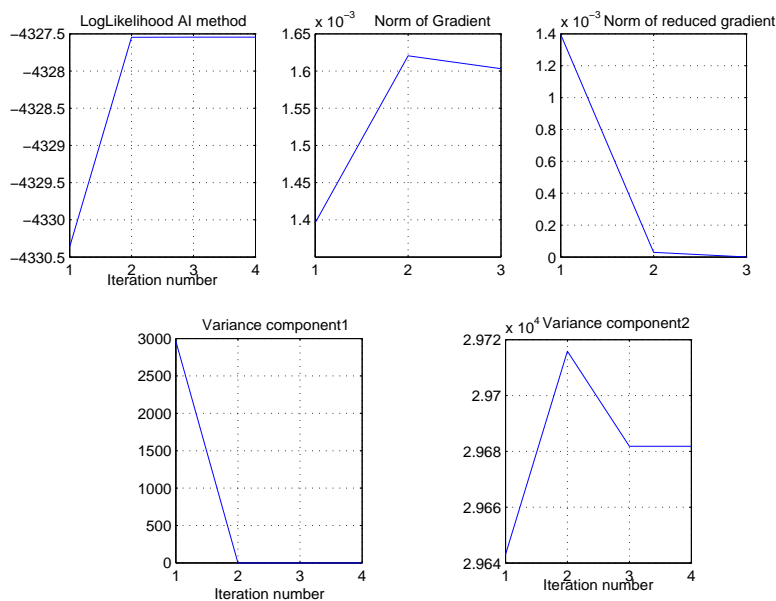}\\
\end{tabular}
\caption{Active-set method with AI approximation for the Hessian,
for data set 2}\label{f8}
\end{figure}

\clearpage
\begin{figure}[h]
\centering
\includegraphics[width=4in]{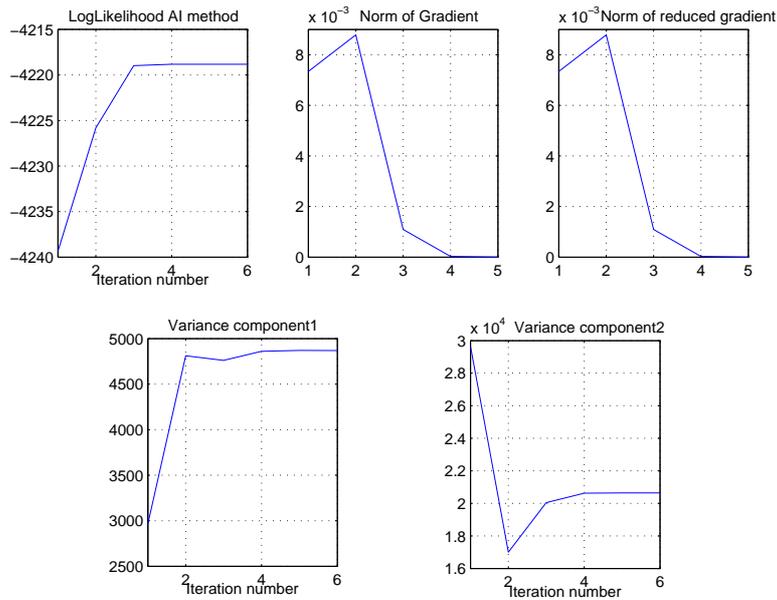}
\caption{The active-set method, AI approximation for the Hessian,
for data set 1}\label{f6}
\end{figure}

\begin{figure}[hf]
\centering
\includegraphics[width=4in]{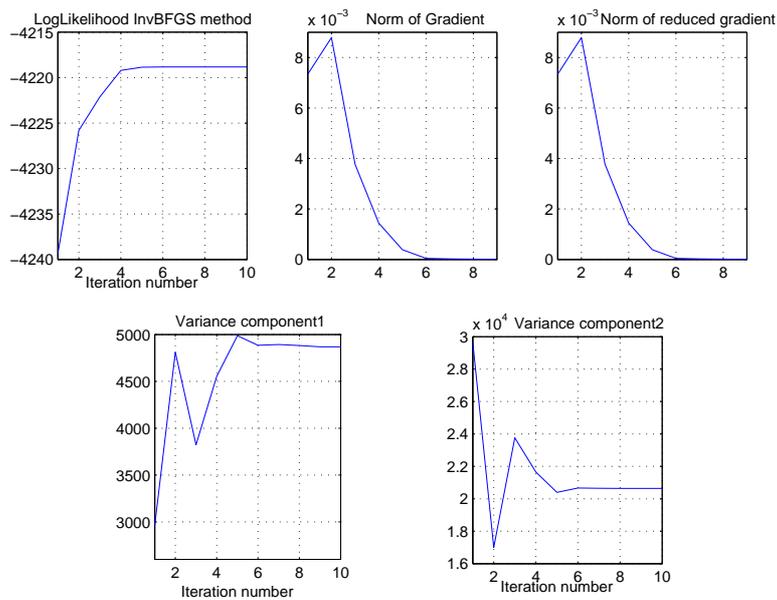}
\caption{The active-set method with quasi-Newton approximation for
the Hessian, for data set 1}\label{f7}
\end{figure}

\clearpage
\section{Conclusions}
In this paper we consider optimization procedures for maximizing the
log-likelihood for REML models used in a QTL mapping setting. We
first show that the standard Newton-AI scheme fails for a problem
where the optimum is located at a constraint boundary. Then we show
how this scheme can be modified to produce a correct solution also
in these cases by including a simple line search procedure. We also
introduce an enhanced quasi-Newton scheme, where the line search
procedure is included and where the average information matrix is
used both as a starting guess and at locations where the curvature
criterion does not hold. A strong side of this method is that for
non-convex functions a better approximation of the Hessian than the
average information matrix can be computed.  Generally, we want to
point out that the unconstrained methods considered in this
framework are sensitive to the choice of the approximation to the
Hessian.

As a second step we describe how an active-set method, which
automatically includes the constraints, can be used for solving the
REML optimization problems. For the data set where the optimum is
located at one of the constraint boundaries, the cpu-time is reduced
by approximately a factor of two compared to the corresponding
unconstrained method.

In our numerical experiments we used the termination criterium
$\|DL\|_{2}^{2} \le 10^{-6}$ or $\|N^{T}DL\|_{2}^{2} \le 10^{-6}$
which turned out to be unnecessary low.

The overall conclusion is that for problems of the type considered
here, the active-set method is robust, and should be preferred
compared to using an unconstrained method. Moreover, the method
using the average information matrix for the Hessian approximation
gives fast and robust results when optimum is located inside the
feasible region.

\bibliographystyle{plain}
\bibliography{ref_LR}

\begin{bibdiv}
\begin{biblist}

\bib{CaHa91}{article}{
      author={Callanan, T.~P.},
      author={Harville, D.~A.},
       title={Some new algorithm for computing restricted maximum likelihood
  estimates of variance components},
        date={1991},
     journal={Journal of Statistical Computation and Simulation},
      volume={38},
       pages={239\ndash 259},
}

\bib{DeLaRu77}{article}{
      author={Dempster, A.~P.},
      author={Laird, N.~M.},
      author={Rubin, D.~B.},
       title={Maximum likelihood from incomplete data via the em algorithm},
        date={1977},
     journal={Journal of the Royal Statistical Society},
      volume={29},
      number={1},
       pages={1\ndash 38},
}

\bib{DrDu06}{misc}{
      author={Druet, T.},
      author={Ducrocq, V.},
       title={Innovations in software packages in quantitative genetics},
        date={2006},
        note={World Congress on Genetics Applied to Livestock Production,Belo
  Horizonte, Brazil},
}

\bib{FeGr89}{article}{
      author={Fernando, R.~L.},
      author={Grossman, M.},
       title={Marker-assisted selection using best linear unbiased prediction},
        date={1989},
     journal={Genet. Sel. Evol.},
      volume={21},
       pages={467\ndash 477},
}

\bib{GiThoCu95}{article}{
      author={Gilmour, A.~R.},
      author={{T}hompson, R.},
      author={{C}ullis, B.R.},
       title={Average information reml: an efficient algorithm for variance
  parameter estimation in linear mixed models},
        date={1995},
     journal={Biometrics},
      volume={51},
       pages={1440\ndash 1450},
}

\bib{Go90}{article}{
      author={Goldgar, D.~E.},
       title={Multipoint analysis of human quantitative genetic variation},
        date={1990},
     journal={American Journal of Human Genetics},
      volume={47},
       pages={957\ndash 967},
}

\bib{HK92}{article}{
      author={Haley, C.~S.},
      author={Knott, S.~A.},
       title={A simple regression method for mapping quantitative trait loci in
  line crosses using flanking markers},
        date={1992},
     journal={Heredity},
      volume={69},
       pages={315\ndash 324},
}

\bib{Ha77}{article}{
      author={Harville, D.~A.},
       title={Maximum likelihood approaches to variance component estimation
  and to related problems},
        date={1977},
     journal={Journal of the American Statistical Association},
      volume={72},
       pages={320\ndash 338},
}

\bib{He63}{inproceedings}{
      author={Henderson, C.~R.},
       title={Selection index and expected genetic advance},
        date={1963},
   booktitle={Statistical genetics and plant breeding},
      editor={Hanson, W.~D.},
      editor={Robinson, H.~F.},
       pages={141\ndash 163},
}

\bib{Meychap23}{book}{
      author={Hill, W.~G.},
      editor={Hill, W.G.},
      editor={McKay, T.F.M.},
       title={Evolution and animal breeding: Reviews on molecular and
  quantitative approaches in honour of a. robertson},
   publisher={Oxford University Press},
        date={1989},
}

\bib{Je97}{article}{
      author={J.~Jensen, P.~Madsen, E. A.~Manysaari},
      author={Thompson, R.},
       title={Residual maximum likelihood estimation of (co)variance components
  in multivariate mixed linear models using average information},
        date={1997},
     journal={Journal of the Indian Society of Agricultural Statistics},
      volume={49},
       pages={215\ndash 236},
}

\bib{JeSa76}{article}{
      author={Jennrich, R.I.},
      author={Sampson, P.F.},
       title={Newton-raphson and related algorithms for maximum likelihood
  variance component estimation},
        date={1976},
     journal={Technometrics},
      volume={18},
       pages={11\ndash 17},
}

\bib{JoTho95}{article}{
      author={Johnson, D.~L.},
      author={Thompson, R.},
       title={Restricted maximum likelihood estimation of variance components
  for univariate animal models using sparse matrix techniques and average
  information},
        date={1995},
     journal={Journal of Dairy Science},
      volume={8},
      number={2},
       pages={449\ndash 456},
}

\bib{LeWe06}{article}{
      author={Lee, S.H.},
      author={van~der Werf, J.H.J.},
       title={An efficient variance component approach implementing an average
  information reml suitable for combined ld linkage mapping with a general
  complex pedigree},
        date={2006},
     journal={Genetics Selection Evolution},
      volume={38},
       pages={25\ndash 43},
}

\bib{LyWa97}{book}{
      author={Lynch, M.},
      author={Walsh, B.},
       title={Genetics and analysis of quantitative traits},
   publisher={{S}inauer {A}ssociates, {I}nc.},
        date={1998},
}

\bib{MeSm96}{article}{
      author={Meyer, K.},
      author={Smith, S.P.},
       title={Restricted maximum likelihood estimation for animal models using
  derivatives of the likelihood},
        date={1996},
     journal={Genetics Selection Evolution},
      volume={28},
       pages={23\ndash 49},
}

\bib{Me89}{article}{
      author={Meyer, K.},
       title={Restricted maximum likelihood to estimate variance components for
  animal models with several random effects using a derivative-free algorithm},
        date={1989},
     journal={Genetics Selection Evolution},
      volume={21},
       pages={317\ndash 340},
}

\bib{NaSo96}{book}{
      author={Nash, S.~G.},
      author={Sofer, A.},
       title={Linear and nonlinear programming},
   publisher={McGraw-Hill International Edition},
        date={1996},
        ISBN={0-07-114537-0},
}

\bib{NeMe64}{article}{
      author={Nelder, J.~A.},
      author={Mead, R.},
       title={A simplex method for function minimization},
        date={1964},
     journal={The Computer Journal},
      volume={7},
       pages={308\ndash 313},
}

\bib{Nocedal}{book}{
      author={Nocedal, J.},
      author={Wright, S.},
       title={Numerical {O}ptimization},
   publisher={Springer-{V}erlag},
     address={New {Y}ork},
        date={1999},
}

\bib{RC07}{article}{
      author={rd, L.~R\"onneg\aa},
      author={Carlborg, \"O.},
       title={Separation of base allele and sampling term effects gives new
  insights in variance component qtl analysis},
        date={2007},
     journal={BMC Genetics},
      volume={8},
      number={1},
}

\bib{ThoMe86}{article}{
      author={{T}hompson, R.},
      author={{M}eyer, K.},
       title={Estimation of variance components: What is missing in the em
  algorithm},
        date={1986},
     journal={Journal of Statistical Computation and Simulation},
      volume={24},
       pages={215\ndash 230},
}

\bib{Tho73}{article}{
      author={Thompson, R.},
       title={The estimation of variance and covariance components with an
  application when records are subject to culling},
        date={1973},
     journal={Biometrics},
      volume={29},
       pages={527\ndash 550},
}

\end{biblist}
\end{bibdiv}
\end{document}